\documentclass[numberedappendix]{emulateapj}
\usepackage{amssymb}

\def\eq#1{\begin{equation} #1 \end{equation}}
\let\mic=\micron

\shorttitle{Temperature inversion in multigrain dust}
\shortauthors{Vinkovi\'{c} D.}

\begin{document}

\title{Temperature inversion on the surface of externally heated
optically thick multigrain dust clouds}

\author{Dejan Vinkovi\'{c}}
\affil{Institute for Advanced Study, School of Natural Sciences,
       Einstein Drive, Princeton, NJ 08540; dejan@ias.edu}

\begin{abstract}
It was recently discovered that the temperature in the surface layer
of externally heated optically thick gray dust clouds increases with
the optical depth for some distance from the surface, as opposed to
the normal decrease in temperature with distance in the rest of the
cloud. This temperature inversion is a result of efficient absorption
of diffuse flux from the cloud interior by the surface dust exposed to
the external radiation.  A micron or bigger size grains experience
this effect when the external flux is of stellar spectrum. We explore
what happens to the effect when dust is a mixture of grain sizes
(multigrain).  Two possible boundary conditions are considered: i) a
constant external flux without constrains on the dust temperature, and
ii) the maximum dust temperature set to the sublimation
temperature. We find that the first condition allows small grains to
completely suppress the temperature inversion of big grains if the
overall opacity is dominated by small grains. The second condition
enables big grains to maintain the inversion even when they are a
minor contributor to the opacity. In reality, the choice of boundary
condition depends on the dust dynamics. When applied to the physics of
protoplanetary disks, the temperature inversion leads to a previously
unrecognized disk structure where optically thin dust can exist inside
the dust destruction radius of an optically thick disk. We conclude
that the transition between the dusty disk and the gaseous inner
clearing is not a sharp edge, but rather a large optically thin
region.
\end{abstract}

\keywords{ accretion, accretion disks --- circumstellar matter --- dust,
extinction --- stars: pre-main-sequence}

\section{Introduction}

Dust is one of the principal components of interstellar and
circumstellar matter. It serves as a very efficient absorber of
starlight, which is dominated by visual and ultraviolet photons, and
reemiter of this absorbed energy into infrared (IR), thereby modifying
the entire spectral energy distribution. Hence, radiative transfer in
dusty environments is an inherent part of the study of star formation,
protoplanetary disk evolution, dusty winds from AGB stars,
circumnuclear environment of AGNs, etc.

In general, radiative transfer is extremely complicated because the
dust in these environments always comes as a mixture of various grain
sizes (multigrain\footnote{Typically, the term {\it multigrain} also
includes all other dust grain properties, like the grain shape and
chemistry. But, for simplicity, we use this term only to describe
grain size effects.}), shapes and chemical compositions. It is,
therefore, common to employ certain approximations that make the
problem computationally manageable. For example, dust grains are
usually approximated with spheres of similar chemical composition, so
that the Mie theory can be easily used for calculating dust cross
sections. Another most commonly used approximation is replacing a
mixture of dust grain sizes by an equivalent (i.e. synthetic or
average) single grain size.  Numerical calculations have shown that
this approximation does not produce a significant change in the
spectral energy distribution \citep{ERR94,Wolf,Carciofi}. It is
assumed that this is particularly appropriate for optically thick dust
clouds.

It has been discovered recently that externally heated optically thick
clouds made of large ($\ga 1\mu$m) single size dust grains produce the
effect of {\it temperature inversion} where the maximum temperature is
within the dust cloud\footnote{Another type of temperature inversion
has been recognized in protoplanetary disks, where additional viscous
heating can increase the disk interior temperature
\citep{Calvet,Malbet}. In contrast, the temperature inversion
discussed here is a pure radiative transfer effect and does not
require any additional assumption (like disk viscosity) to operate.},
at the visual optical depth of $\tau_V\sim$1, instead of on the very
surface exposed to the stellar heating \citep{Isella,VIJE06}. The
cause of this inversion within the surface layer is its ability to
efficiently absorb the diffuse IR radiation originating from the
cloud's interior - a process similar to the ``greenhouse effect.''

It is not known, however, if multigrain dust would produce the same
temperature inversion effect. Here we employ analytical multigrain
radiative transfer to study the surface of optically thick clouds.
Two possible types of boundary conditions are explored: i) a constant
external flux heating the cloud, with no limits on the dust
temperature, and ii) a fixed maximum dust temperature corresponding to
dust sublimation. In \S\ref{single_grain} we describe the analytic
method and apply it on single size dust. In the next section
\S\ref{multigrain} we apply the method on multigrain dust. After that
in \S\ref{transverse} we explore a possibility of surface thermal
cooling in transverse direction.  Some aspects of our result are
discussed in \S\ref{discussion}, with our conclusion in
\S\ref{conclusion}.

\section{Optically thick single size grain dust cloud}
\label{single_grain}

In order to understand effects of multigrain dust on the temperature
structure of optically thick clouds, we first have to take a look at
the single size dust grain clouds. Here we reiterate analytic
techniques described in the literature \citep{Isella,VIJE06} and then
expand our approach to multigrain dust in the following sections.

\begin{figure}
\epsscale{1.15}
 \plotone{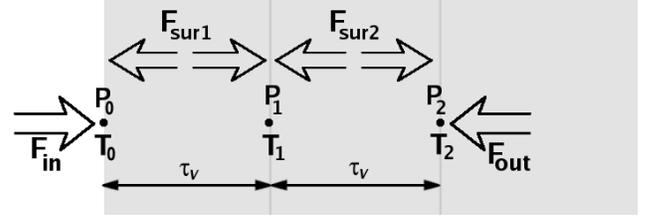}
 \caption{\label{Sketch1} Sketch of the surface of an optically thick
single size grain dust cloud illuminated from outside.  See
\S\ref{single_grain} for details.}
\end{figure}

Consider dust grains at three different locations (Figure \ref{Sketch1}):
on the very surface (point {\bf P$_0$}), at optical depth $\tau_V$
(point {\bf P$_1$}) and at optical depth $2\tau_V$ (point {\bf
P$_2$}). The optical depth $\tau_V$ is defined at the peak wavelength
of the bolometric temperature of external flux $F_{in}$ illuminating
the cloud. We will be interested in cases where dust can reach
sublimation temperatures (between $\sim$1,000K and 2,000K for
interstellar dust) and where the external flux source is a star-like
object peaking its emission at submicron wavelengths. For the purpose
of this paper we use the optical depth $\tau_V$ in visual, but one can
adjust it to the required wavelength of interest.

The surface layer between points {\bf P$_0$} and {\bf P$_1$} emits the
thermal flux $F_{sur1}$. We assume that it emits equally on both
sides (toward the left and right in Figure \ref{Sketch1}). This
approximation is valid if the layer is optically thin at the peak
wavelength of its IR emission. If we define $q=\sigma_V/\sigma_{IR}$,
where $\sigma_V$ and $\sigma_{IR}$ are dust absorption cross sections in
visual and IR, then the IR optical depth requirement is
 \eq{\label{tauIR}
   \tau_{IR}={\tau_V\over q}\la 1.
 }
The same requirement holds for the second surface layer between points
{\bf P$_1$} and {\bf P$_2$} and its thermal flux $F_{sur2}$.
Each of these two layers attenuates the external flux by
$exp(-\tau_V)$ and the thermal fluxes by
$exp(-\tau_{IR})=exp(-\tau_V/q)$.

Since the cloud is optically thick to its own radiation, no net
flux can go through the cloud. The total flux coming from the left
at any point in the cloud is equal to the total flux coming from
the right. This yields flux balance equations at points {\bf
P$_0$}, {\bf P$_1$} and {\bf P$_2$}
 \eq{\label{F0}
  F_{in} = F_{sur1} + F_{sur2}\, e^{-\tau_V/q} + F_{out}\, e^{-2\tau_V/q}
 }
 \eq{\label{F1}
  F_{in}\, e^{-\tau_V}+ F_{sur1} = F_{sur2} + F_{out}\, e^{-\tau_V/q}
 }
 \eq{\label{F2}
  F_{in}\, e^{-2\tau_V} + F_{sur1}\, e^{-\tau_V/q} + F_{sur2} = F_{out}
 }
where $F_{out}$ is the thermal flux coming out of the cloud interior
at {\bf P$_2$}.

We want the external flux $F_{in}$ at point {\bf P$_2$} to be attenuated enough
to make diffuse flux the dominant source of heating.
This requirement means that $2\tau_V\ga 1$, which in combination with
equation \ref{tauIR} gives the allowed range for $\tau_V$
 \eq{
   0.5\la \tau_V \la q.
 }
From equations \ref{F0}-\ref{F2} we further get
 \eq{\label{Fsur2}
  F_{sur2}=F_{in} {1+e^{-\tau_V} \over 1+e^{-\tau_V/q}} - F_{out}e^{-\tau_V/q}
 }
 \eq{\label{Fsur1}
  F_{sur1}=F_{sur2} - F_{in} e^{-\tau_V} {1+e^{-\tau_V} \over 1+e^{-\tau_V/q}}.
 }

A dust grain at {\bf P$_0$} absorbs $\sim \sigma_V F_{in}$ of the
external directional flux. It also absorbs $\sim 2\sigma_{IR}F_{IR}$
of any infrared diffuse flux $F_{IR}$, where the factor 2 accounts
for absorption from 2$\pi$ sr. The grain emits at its temperature
$T_0$ into 4$\pi$ sr, so that the energy balance is
 \[
  \sigma_VF_{in} + 2\sigma_{IR}(F_{sur1} + F_{sur2}\, e^{-\tau_V/q}
     + F_{out}\, e^{-2\tau_V/q}) =
 \]
 \eq{
     = 4\sigma_{IR}\sigma_{SB}T_0^4
 }
where $\sigma_{SB}$ is the Stefan-Boltzmann constant. Using the flux balance
in equation \ref{F0} and $q=\sigma_V/\sigma_{IR}$, we get
 \eq{\label{Fin}
   F_{in}={4\sigma_{SB}\over q+2} T_0^4.
 }

Similarly, we can write the energy balance for a dust grain at {\bf
P$_1$} and {\bf P$_2$}
 \eq{\label{ebT1}
  qF_{in}\, e^{-\tau_V}+ 2F_{sur1} + 2F_{sur2} + 2F_{out}\, e^{-\tau_V/q}
    = 4\sigma_{SB}T_1^4
 }
 \eq{\label{ebT2}
  qF_{in}\, e^{-2\tau_V} + 2F_{sur1}\, e^{-\tau_V/q} + 2F_{sur2} + 2F_{out}.
    = 4\sigma_{SB}T_2^4
 }

We approximate the interior flux as
 \eq{\label{Fout}
  F_{out} \sim \sigma_{SB}T_2^4.
 }
This is a good approximation of the interior for gray dust and an
overestimate for non-gray dust, where temperature decreases
with optical depth.
Now we can continue deriving temperatures in two possible ways.

{\it METHOD 1}: Combining equations \ref{Fsur2},\ref{Fsur1},
\ref{Fin}, \ref{ebT1}, \ref{ebT2} and \ref{Fout} yields
 \eq{\label{T1_method1}
  T_1^4 = {A+B \over q+2} T_0^4
 }
 \eq{\label{T2_method1}
  T_2^4 = {2B \over q+2} T_0^4
 }
 \eq{\label{A_method1}
  A = qe^{-\tau_V}+2(1-e^{-\tau_V}){1+e^{-\tau_V}\over 1+e^{-\tau_V/q}}
 }
 \eq{\label{B_method1}
  B = qe^{-2\tau_V}+2(1-e^{-\tau_V-\tau_V/q}){1+e^{-\tau_V}\over 1+e^{-\tau_V/q}}.
 }

\begin{figure}
\epsscale{1.15}
 \plotone{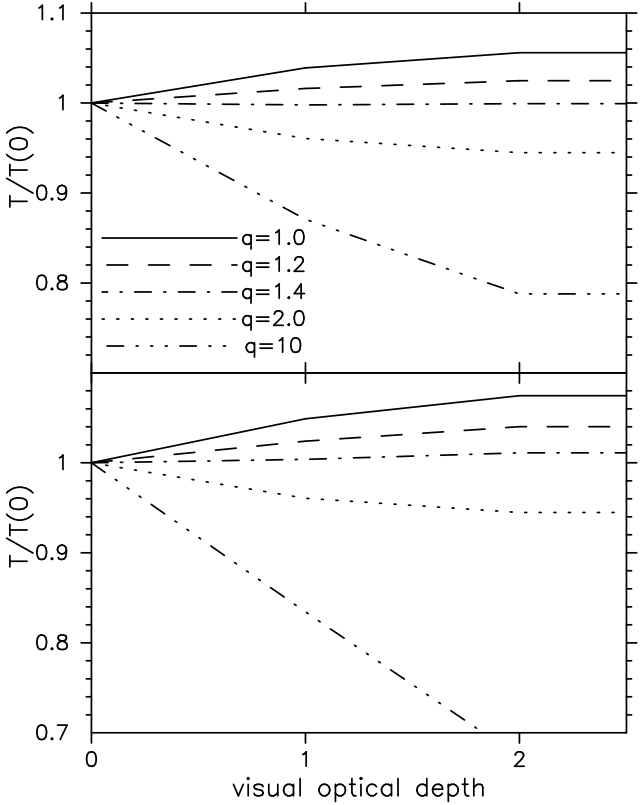}
 \caption{\label{Temperature1}
Surface temperature of an optically thick single size grain dust cloud for various
visual to infrared opacity ratios $q$.
The upper panel shows the solution based on equations \ref{T1_method1} and \ref{T2_method1}
(method 1). The lower panel is based on equations \ref{T2_method2} and \ref{T1_method2}
(method 2). Optical depth step is $\tau_V=1$. Both solutions show temperature inversion
(increase of temperature with optical depth) for small $q$.
 }
\end{figure}

The upper panel of Figure \ref{Temperature1} shows this temperature profile for $\tau_V=1$. The
effect of temperature inversion $T_2>T_0$ is present when
\hbox{$q<q_{limit}\sim 1.4$}. Only grains larger than about one micron
can have such a small value of $q$, which makes them almost ``gray''
in the near IR. This inversion does not appear in non-gray dust
($q>q_{limit}$) where the temperature decreases monotonically with
distance from the cloud surface.

{\it METHOD 2}: We can use equations \ref{F0}-\ref{F2} to express
$F_{out}$ as a function of $F_{in}$ and then use equations \ref{Fin} and
\ref{Fout} to obtain
 \eq{\label{T2_method2}
  T_2^4 = {4\over q+2}\cdot {1+e^{-\tau_V}(1+e^{-\tau_V}-e^{-\tau_V/q}).
           \over 1+e^{-\tau_V/q}} T_0^4
 }
Similarly, we can express $F_{sur1}$ and $F_{sur2}$ as a function of $F_{in}$
and then use equation \ref{ebT1} to obtain
 \eq{\label{T1_method2}
  T_1^4 = {1\over q+2}\left[ qe^{-\tau_V} +
           2{2+e^{-\tau_V}-e^{-\tau_V - \tau_V/q} \over 1+e^{-\tau_V/q}} \right]T_0^4.
 }
The lower panel of Figure \ref{Temperature1} shows this result for $\tau_V=1$.
It differs slightly from previous solution in Method 1, but gives qualitatively the same
result. The temperature inversion now exists for
\hbox{$q<q_{limit}\sim 1.5$}. Henceforth, we use Method 1 for further analysis.

\section{Optically thick multigrain cloud}
\label{multigrain}

\subsection{Constant external flux as the boundary condition}
\label{multigrain_const_flux}

We consider now an externally heated optically thick dust cloud made
of $N$ grain sizes, under the condition of constant external flux and
no limits on the dust temperature. A dust grain of the $i^{th}$ size
at a point of $\tau_V$ optical distance from the cloud surface is
heated by the attenuated stellar flux $F_{in}exp(-\tau_V)$, by the
local diffuse flux $F_{sur}$ coming from the direction of the surface,
and by the local diffuse flux $F_{out}$ coming out of the cloud
interior. The thermodynamic equilibrium of the dust grain gives
 \eq{\label{ebi}
 \sigma_{V,i}\, F_{in}e^{-\tau_V} +2\sigma_{IR,i}\, (F_{sur} +
      F_{out})  = 4\sigma_{IR,i}\sigma_{SB}\, T_i^4(\tau_V).
 }
Factors 2 and 4 account for absorption from 2$\pi$ sr and emission
into 4$\pi$ sr.

On the very surface of the cloud $\tau_V=0$ and $F_{sur}=0$, therefore
$F_{in}=F_{out}$, which yields
 \eq{\label{Fini}
    T_i^4(0) = (q_i+2){F_{in}\over 4\sigma_{SB}}
 }
where $q_i=\sigma_{V,i}/\sigma_{IR,i}$.  This shows that {\it dust
grains of different sizes have different temperatures on the surface
of an optically thick cloud}.  The lowest temperature among the grain
sizes is acquired by the largest grains because they have the smallest
$q_i$.

On the other hand, the contribution of the external flux is negligible
$F_{in}exp(-\tau_V)\to 0$ when $\tau_V\gg 1$, therefore
$F_{sur}=F_{out}$, which yields {\it the same temperature for all grain sizes}
 \eq{\label{T_tauV_gg_1}
   T_i^4(\tau_V\gg 1) =  F_{out}/\sigma_{SB}.
 }

A general relationship between grain temperatures is derived by
subtracting equation \ref{ebi} for a grain $i$ from the same equation
for a grain $j$ and then use equation \ref{Fini} to obtain
 \eq{\label{Tij}
  T_j^4(\tau_V) = T_i^4(\tau_V) + {q_j-q_i\over q_i+2}e^{-\tau_V} T_i^4(0)
 }
Since $q$ scales inversely with the grain size, equation
\ref{Tij} shows that {\it smaller grains always have a higher
temperature than bigger grains at any point in the cloud}, with
the limit $T_j \sim T_i$ when $exp(-\tau_V)\ll 1$.

Optical depth is now a cumulative contribution of all grain sizes in the mix.
 \eq{
   \tau_\lambda \propto \sum\limits_{i=1}^N n_i\sigma_{\lambda,i},
 }
where $n_i$ is the number density of the i$^{th}$ grain size.
If we scale optical depth relative to $\tau_V$ then
 \eq{
   \tau_{IR} = \tau_V { \sum_{i=1}^N n_i\sigma_{IR,i}\over
                        \sum_{i=1}^N n_i\sigma_{V,i} } =
               \tau_V \sum\limits_{i=1}^N {\Upsilon_{V,i}\over q_i},
 }
where $\Upsilon_{\lambda,i}$ is the relative contribution of the $i^{th}$ grain
to the dust opacity at wavelength $\lambda$
 \eq{
   \Upsilon_{\lambda,i} = { n_i\sigma_{\lambda,i}\over
                        \sum_{j=1}^N n_j\sigma_{\lambda,j} }.
 }

\begin{figure}
\epsscale{1.15}
 \plotone{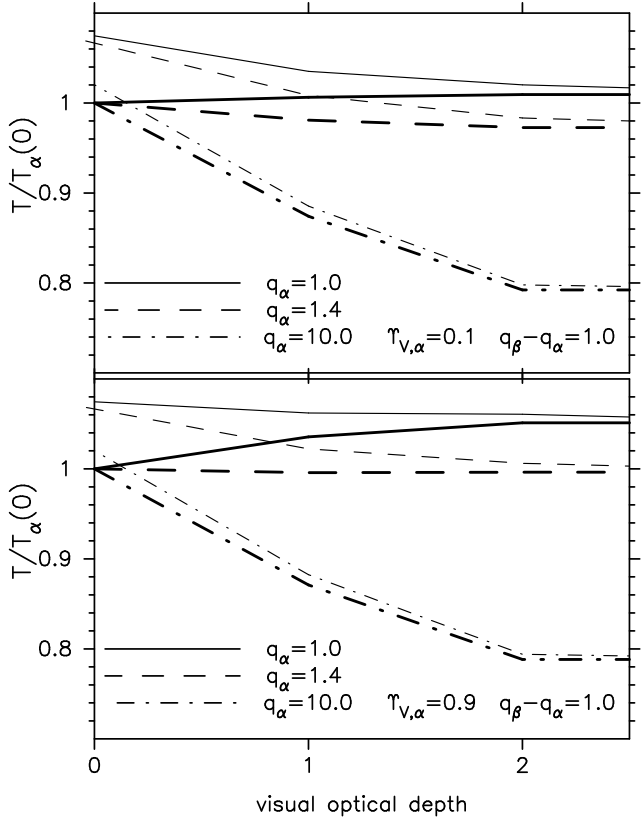}
 \caption{\label{Temp_multigrain_whole_fig1}
Temperature of two types of grains in the surface of an optically
thick multigrain dust cloud: larger grains $\alpha$ ({\it thick
lines}, equations \ref{T1alpha_whole} and \ref{T2alpha_whole}) and
smaller grains $\beta$ ({\it thin lines}, equation \ref{Tsmall_n}).
Line styles correspond to different ratios $q_\alpha$
of visual to infrared opacity of big grains. The same ratio for small
grains is fixed to $q_\beta=q_\alpha+1$.  Panels show results for two
relative contributions of big grains to the visual opacity:
$\Upsilon_{V\alpha}=0.1$ in the upper panel and
$\Upsilon_{V\alpha}=0.9$ in the lower panel.  Optical depth step is
$\tau_V=1$.
Notice how the temperature inversion in big grains (small $q_\alpha$)
disappears when small grains dominate the overall opacity (upper
panel). The inversion exists only if bigger grains dominate (lower
panel).
 }
\end{figure}

\begin{figure}
\epsscale{1.15}
 \plotone{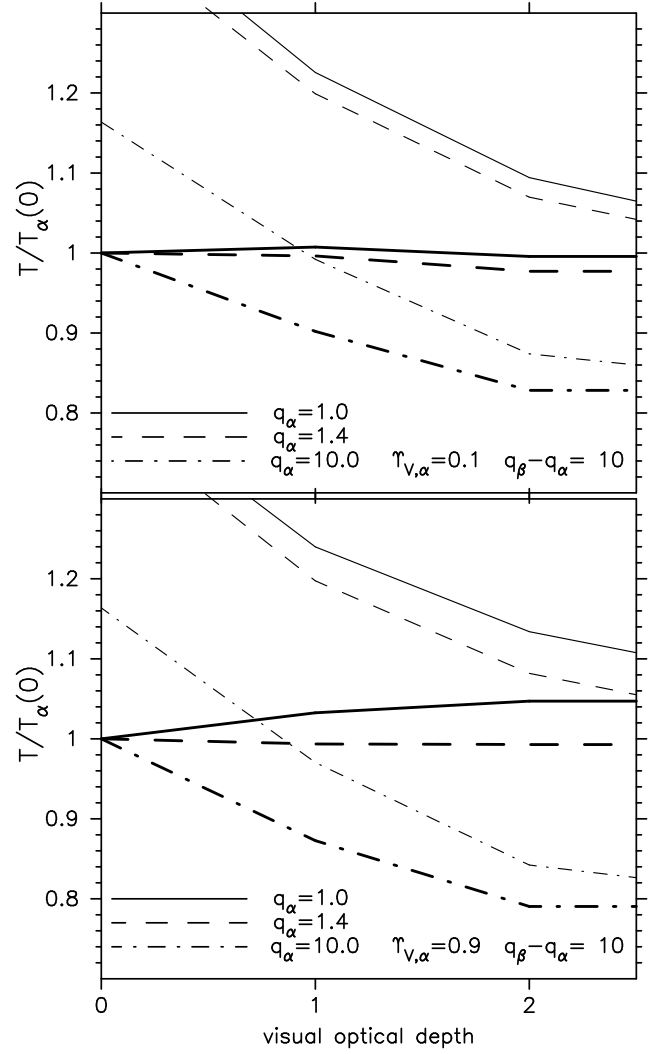}
 \caption{\label{Temp_multigrain_whole_fig2}
Same as Fig. \ref{Temp_multigrain_whole_fig1}, but for $q_\beta=q_\alpha+10$.
 }
\end{figure}

Consider now a model similar to figure \ref{Sketch1}, except that the dust
is multigrain.  For simplicity and clarity of the following
analysis, we work with only two grain sizes: a ``big'' grain {\it
$\alpha$} with $q_\alpha\sim 1$ and a ``small'' grain {\it $\beta$}
with $q_\beta>q_\alpha$.  The infrared optical depth step is
 \eq{\label{tauIR_multigrain}
  \tau_{IR}=\tau_V\left({\Upsilon_{V,\alpha}\over q_\alpha} +
                        {\Upsilon_{V,\beta}\over q_\beta} \right).
 }
According to equation \ref{Tij} the temperature of small grains at
optical depth $k\tau_V$ (point {\bf P$_k$}) is
 \eq{\label{Tsmall_n}
  T_{k,\beta}^4 = T_{k,\alpha}^4 + {q_\beta-q_\alpha\over q_\alpha+2}e^{-k\tau_V}
                                   T_{0,\alpha}^4.
 }
The temperature of big grains at point {\bf P$_0$} is (equation \ref{Fini})
 \eq{\label{T_0alpha}
  T_{0,\alpha}^4 = (q_\alpha+2){F_{in}\over 4\sigma_{SB}},
 }
while for the other temperatures we need the flux balance at points
{\bf P$_0$}, {\bf P$_1$} and {\bf P$_2$}. Since the balance  is
the same as in equations \ref{F0}-\ref{F2}, except that the infrared
step $\tau_V/q$ is replaced by $\tau_{IR}$, we use the procedure described in
\S\ref{single_grain} and derive
 \eq{\label{T1alpha_}
  T_{1,\alpha}^4 = {A_\alpha \over q_\alpha+2}T_{0,\alpha}^4 +
                   {F_{out}\over 2\sigma_{SB}}
 }
 \eq{\label{T2alpha_}
   T_{2,\alpha}^4 = {B_\alpha \over q_\alpha+2}T_{0,\alpha}^4 +
                   {F_{out}\over 2\sigma_{SB}}
 }
 \eq{
  A_\alpha = q_\alpha\, e^{-\tau_V}+
             2\,(1+e^{-\tau_V}){1+e^{-\tau_V}\over 1+e^{-\tau_{IR}}}
 }
 \eq{
  B_\alpha = q_\alpha\, e^{-2\tau_V}+
             2\,(1+e^{-\tau_V-\tau_{IR}}){1+e^{-\tau_V}\over 1+e^{-\tau_{IR}}}.
 }

The interior flux $F_{out}$ is now a cumulative contribution of all
grain sizes according to their relative contribution to the dust
opacity. In our two-size example
 \eq{\label{Fout_multi}
  F_{out}/\sigma_{SB}=\Upsilon_{IR,\alpha} T_{2,\alpha}^4 +
                      \Upsilon_{IR,\beta} T_{2,\beta}^4.
 }
Combined with equation \ref{Tsmall_n} gives
 \eq{\label{Foutc}
  F_{out}/\sigma_{SB}=T_{2,\alpha}^4 + {C_\alpha\over q_\alpha+2}T_{0,\alpha}^4
 }
 \eq{\label{calpha}
   C_\alpha = \Upsilon_{IR,\beta}(q_\beta - q_\alpha)e^{-2\tau_V}.
 }

Putting together equations \ref{T1alpha_}, \ref{T2alpha_} and
\ref{Foutc} yields the solution
 \eq{\label{T1alpha_whole}
  T_{1,\alpha}^4 = {A_\alpha+ B_\alpha+C_\alpha\over q_\alpha+2}T_{0,\alpha}^4
 }
 \eq{\label{T2alpha_whole}
  T_{2,\alpha}^4 = {2B_\alpha+C_\alpha\over q_\alpha+2}T_{0,\alpha}^4.
 }

The resulting temperature is plotted in Figures
\ref{Temp_multigrain_whole_fig1} and \ref{Temp_multigrain_whole_fig2}.
The upper panel in each figure shows the result when small grains
dominate the opacity ($\Upsilon_{V,\alpha}=0.1$). The temperature
inversion in big grains is suppressed because the local diffuse flux
is dictated by small grains. If big grains dominate the opacity (lower
panels, $\Upsilon_{V,\alpha}=0.9$) then the temperature inversion is
preserved.

\subsection{Sublimation temperature as the boundary condition}
\label{multigrain2}

\begin{figure}
\epsscale{1.15}
 \plotone{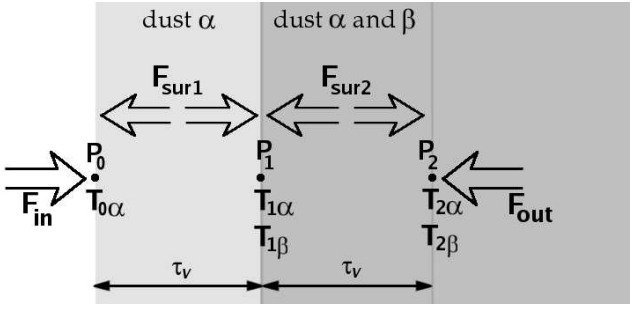}
 \caption{\label{Sketch2}
Sketch of the surface of an optically thick multigrain dust cloud illuminated from outside.
 See \S\ref{multigrain2} for details.
 }
\end{figure}

Now we consider a dust cloud hot enough on its illuminated surface to
sublimate dust grains warmer than the sublimation temperature
$T_{sub}$. The flux entering the cloud is adjustable to accommodate
any temperature boundary condition. From equation \ref{Fini} we see
that small grains are the first to be removed from the immediate
surface. If the external flux is high enough then the immediate
surface is populated only by $q_i\sim 1$ grains (``big grains''). All
other grains (``small grains'') would survive somewhere within the
cloud, at a distance where the local flux is reddened enough by big
grains to be absorbed less efficiently.

\begin{figure}
\epsscale{1.15}
 \plotone{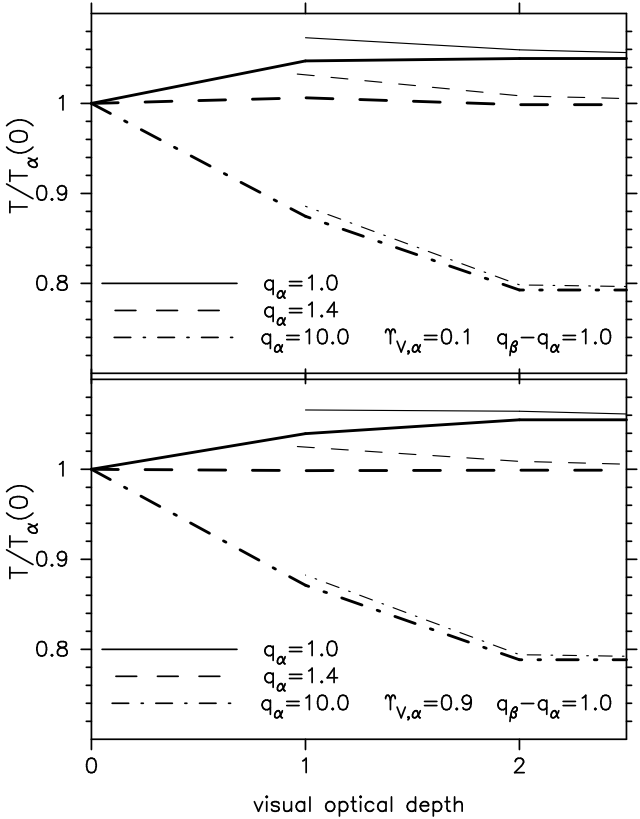}
 \caption{\label{Temp_multigrain_fig1}
Same as Figure \ref{Temp_multigrain_whole_fig1}, except that small
grains are removed from the immediate surface by sublimation (Figure
\ref{Sketch2}). The temperature of larger grains is calculated from
equations \ref{T1alpha} and \ref{T2alpha}.  Despite being mixed with
smaller grains, larger grains show temperature inversion even when
smaller grains dominate the opacity.
}
\end{figure}

\begin{figure}
\epsscale{1.15}
 \plotone{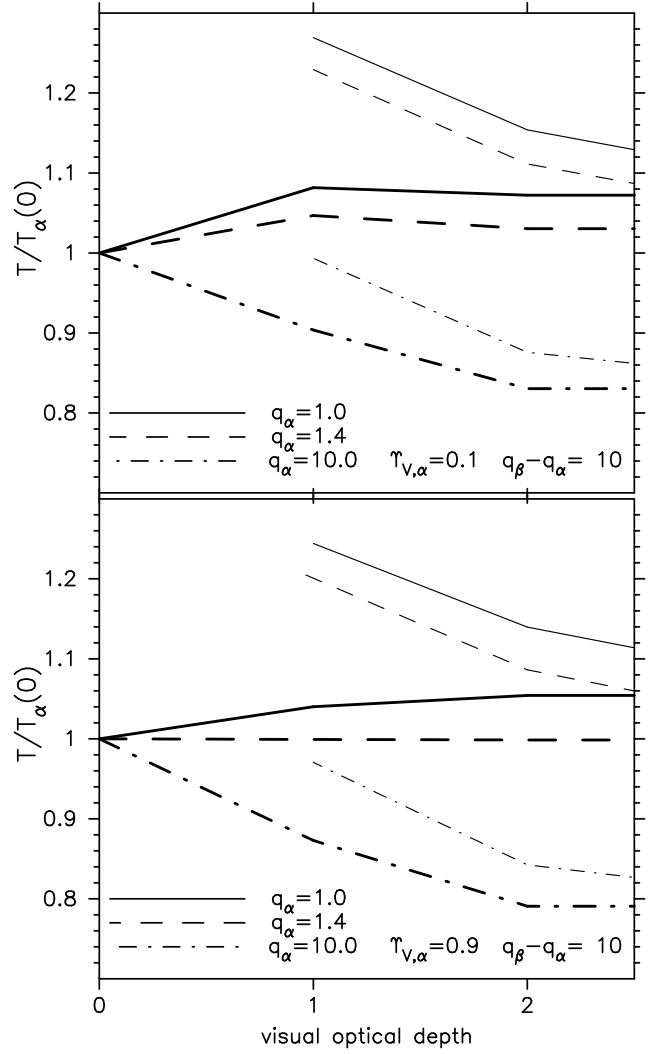}
 \caption{\label{Temp_multigrain_fig2}
Same as Figure \ref{Temp_multigrain_fig1}, but for $q_\beta=q_\alpha+10$.
 }
\end{figure}

We again apply our two-size dust model, except that the surface layer
of visual optical depth $\tau_V$ is occupied only by big grains (see
Figure \ref{Sketch2}). It is too hot for small grains to survive
within this layer. Both grains exist at optical distances larger than
$\tau_V$ from the surface.
Following the same procedure as in \S\ref{single_grain}, we write
the flux balance at points {\bf P$_0$}, {\bf P$_1$} and {\bf P$_2$} (see
Figure \ref{Sketch2})
 \eq{\label{F0multi}
  F_{in} = F_{sur1} + F_{sur2}\, e^{-\tau_V/q_\alpha} + F_{out}\, e^{-\tau_V/q_\alpha-\tau_{IR}}
 }
 \eq{\label{F1multi}
  F_{in}\, e^{-\tau_V}+ F_{sur1} = F_{sur2} + F_{out}\, e^{-\tau_{IR}}
 }
 \eq{\label{F2multi}
  F_{in}\, e^{-2\tau_V} + F_{sur1}\, e^{-\tau_{IR}} + F_{sur2} = F_{out}
 }
where the IR optical depth between points {\bf P$_1$} and {\bf P$_2$}
is given in equation \ref{tauIR_multigrain}.

Small grains do not exist now at point {\bf P$_0$}, but their
temperature at other points is still described by equation
\ref{Tsmall_n}.  Deriving big grain temperatures at {\bf
P$_1$} and {\bf P$_2$} is now a straightforward procedure already
described in \S\ref{single_grain} and \S\ref{multigrain_const_flux}
 \eq{
  A_\alpha^\prime = q_\alpha\, e^{-\tau_V}+2\, {1+e^{-\tau_V}\over 1+e^{-\tau_V/q_\alpha}}
                   -2\, e^{-\tau_V}{1+e^{-\tau_V}\over 1+e^{-\tau_{IR}}}.
 }
 \eq{
  B_\alpha^\prime = q_\alpha\, e^{-2\tau_V}+2\, {1+e^{-\tau_V}\over 1+e^{-\tau_V/q_\alpha}}
                   -2\, e^{-\tau_V-\tau_{IR}}{1+e^{-\tau_V}\over 1+e^{-\tau_{IR}}}
 }
 \eq{\label{T1alpha}
  T_{1,\alpha}^4 = {A_\alpha^\prime+ B_\alpha^\prime+C_\alpha\over q_\alpha+2}T_{0,\alpha}^4
 }
 \eq{\label{T2alpha}
  T_{2,\alpha}^4 = {2B_\alpha^\prime+C_\alpha\over q_\alpha+2}T_{0,\alpha}^4.
 }
The resulting temperature is plotted in Figures \ref{Temp_multigrain_fig1}
and \ref{Temp_multigrain_fig2}. Two
important results are deduced for big grains from this solution:
i) big grains maintain the temperature inversion and ii) if small
grains set the maximum temperature limit then big grains, which are always
colder than small grains (see equation \ref{Tij}), cannot reach
the sublimation temperature.

However, since we do not put limits on the external flux, the
solution that allows the maximum possible external flux is the one
that also maximizes $F_{out}$. From equation \ref{Fout_multi} we see
that the maximum $F_{out}=\sigma_{SB}T_{sub}$ is achieved when
$T_{2,\alpha}=T_{2,\beta}=T_{sub}$. Such a solution is not possible
unless $\Upsilon_{IR,\beta}=0$. Therefore, {\it dust temperatures
are maximized when all small grains sublimate away from the
surface region and exist only at optical depths of
$exp(-\tau_V)\ll 1$. In other words, the external flux is
maximized when the cloud surface "belongs" exclusively to big
grains only}.

\section{Transverse cooling of the cloud surface}
\label{transverse}

Solutions presented so far assume no time variability in any of the
model parameters. The external flux is adjusted by hand to the value
that maximizes the external flux and keeps dust temperatures below
sublimation. In reality, however, this is a dynamical process where the
equilibrium is established by dust moving around and sublimating
whenever its temperature exceeds the sublimation point.

The existence of temperature inversion is difficult to understand
under such dynamical conditions. Since the very surface of the
cloud is {\it below} the sublimation point, its dust can survive
closer to the external energy source than the dust within the
cloud. Therefore, the distance of the cloud from the source is not
defined by the very surface, but rather by the dust at the peak
temperature {\it within} the cloud. Imagine now that the whole
cloud is moving closer to the source. From the cloud point of
reference, this dynamical process is equivalent to increasing the
external flux. According to solutions in \S\ref{single_grain} and
\S\ref{multigrain}, the peak temperature within the cloud will
exceed sublimation at a certain distance from the source and the
dust will start to sublimate. This point is the distance at which
the dust cloud seemingly stops. However, there is nothing to stop
the very surface layer of the cloud to move even closer than the
rest of the cloud because its temperature is below sublimation.

Notice that we can not resolve this issue in the approximation of an
infinite dusty slab because the transverse optical depth (parallel to
the slab surface) is always infinite. No flux can escape the slab in
the transverse direction. Hence, the radiative transfer in a slab does
not depend on spatial scale and only the optical depth matters. The
spatial extension of the surface layer is irrelevant and solutions
from \S\ref{single_grain} and \S\ref{multigrain} are applicable
irrespectively of the dust dynamics on the spatial scale.

In reality, however, the cloud is finite and dust sublimation can
eventually make the cloud optically thin in the transverse
direction. This optical depth gap enables thermal radiation to flow
out in the transverse direction and thus provides a channel for
thermal cooling. Under such conditions the surface layer cam move
closer and closer to the energy source by simultaneously expanding the
size of the transverse optical depth gap, which then increases the
amount of escaping thermal flux. Since the very surface dust is
getting closer to the energy source, its temperature increases and
eventually it reaches the dust sublimation. At that moment the
radiative and dynamical equilibria are established and the whole dust
cloud seemingly stops.

\begin{figure}
\epsscale{1.15}
 \plotone{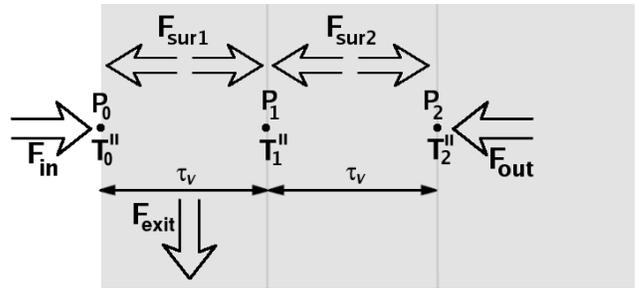}
 \caption{\label{Sketch3}
Sketch of the surface of an optically thick dust cloud illuminated from outside and
cooled through the surface layer in the transverse direction.
 See \S\ref{transverse} for details.
 }
\end{figure}

The resulting dust cloud has a large surface zone where the
thermal flux can transversely escape the cloud. Here we make an
attempt to describe this on a quantitative level through a model
shown in Figure \ref{Sketch3}. It is similar to the infinite slab model
described in \S\ref{single_grain} except that now we introduce the
flux $F_{exit}$ exiting the cloud transversely through the surface
layer.

The flux balance equations \ref{F1} and \ref{F2} remain unchanged,
while equation \ref{F0} is changed to
 \eq{\label{F0_exit}
  F_{in} = F_{exit} + F_{sur1} + F_{sur2}\, e^{-\tau_V/q} + F_{out}\,
  e^{-2\tau_V/q}.
 }
By following the same procedure as in \S\ref{single_grain}, we
obtain temperatures $T^{\prime\prime}_k$ at points {\bf P$_k$}
 \eq{\label{Fin_thin}
   F_{in}={4\sigma_{SB}\over q+2} T^{\prime\prime 4}_0 +{2\over q+2} F_{exit}
 }
\eq{\label{T1_thin}
  T^{\prime\prime 4}_1 = {A+B \over q+2} T^{\prime\prime 4}_0 -\left({1\over 1+e^{-\tau_V/q}}-{A+B\over 2(q+2)}\right)F_{exit}
 }
 \eq{\label{T2_thin}
  T^{\prime\prime 4}_2 = {2B \over q+2} T^{\prime\prime 4}_0 -\left({1\over 1+e^{-\tau_V/q}}-{B\over q+2}\right)F_{exit}
 }
where A and B are given in equations \ref{A_method1} and
\ref{B_method1}. The values in parentheses are always positive.

\begin{figure}
\epsscale{1.15}
 \plotone{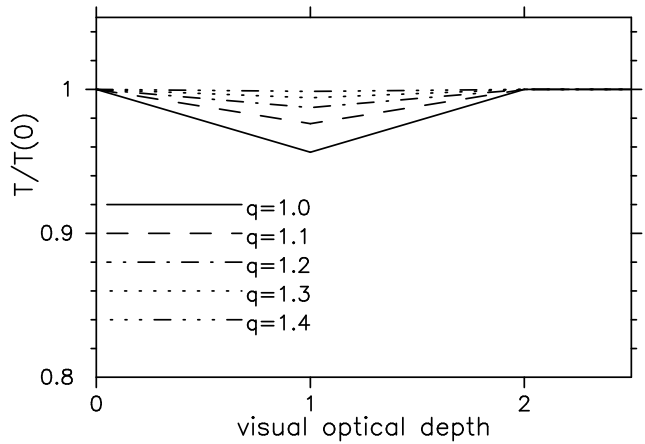}
 \caption{\label{Temperature_exit}
Surface temperature of an optically thick single size grain dust cloud
cooled transversely through the surface layer (equations \ref{T1_thin}
and \ref{T2_thin}).  The amount of transverse cooling is adjusted to
the maximum temperature $T(0)$ at the very surface (equation
\ref{Fexit_min}).  Line styles correspond to different ratios $q$ of
visual to infrared opacity.  Optical depth step is $\tau_V=1$.
Increase in $T(0)$, combined with the temperature inversion (see also
Fig. \ref{Temperature1}), creates a local temperature minimum within
the surface layer.
 }
\end{figure}

We compare this with the original solution in equations \ref{Fin},
\ref{T1_method1} and \ref{T2_method1}. Notice that keeping the
interior temperature to be the same as before
($T^{\prime\prime}_2=T_2$) requires a larger surface temperature
($T^{\prime\prime}_0>T_0$). If the surface dust of
$T^{\prime\prime}_0$ is dynamically driven toward the external
energy source, as we discussed above, then it will eventually
reach the sublimation temperature $T_{sub}$. This dynamical
process is accompanied by dust sublimation in the cloud interior.
Sublimation creates a transverse optical depth gap and maintains
$T^{\prime\prime}_2=T_{sub}$. The established equilibrium
condition $T^{\prime\prime}_0=T^{\prime\prime}_2=T_{sub}$ yields
 \eq{\label{Fexit_min}
 F_{exit} = {2B-q-2 \over
 (q+2)/(1+e^{-\tau_V/q})-B}\,\,\,T^4_{sub}.
  }
This flux is positive for values of $q$ that produce temperature
inversion ($q<q_{limit}$). It originates from the
diffuse radiation trapped within the cloud. 
Figure \ref{Temperature_exit} shows temperature profiles for
$q<q_{limit}$ based on $F_{exit}$ from equation \ref{Fexit_min}.
The surface layer now has a local temperature minimum at
$T^{\prime\prime}_1$.

\section{Discussion}
\label{discussion}

According to equation \ref{T_tauV_gg_1}, the temperature of dust
optically deep inside an optically thick cloud depends solely on
the local diffuse flux. Since this can give a misleading
impression that temperature effects on the cloud surface have no
influence on the cloud interior, it is worth explaining why the
surface solution is so important for the overall solution.

At optical depths $\tau_V\ga 2$ the flux is exclusively diffuse and
the dust temperature is dictated by equation
\ref{T_tauV_gg_1}. Notice, however, that even though we set the limit
of zero net flux flowing through the cloud, there is no limit on the
absolute value of the local diffuse flux. Arbitrarily large but equal
diffuse fluxes can flow in opposite directions producing the net zero
flux. It is the radiative transfer in the surface layer that sets the
absolute scale. The cloud surface of $\tau_V\la 2$ converts the
directional external flux of stellar spectral shape into the diffuse
thermal flux. Hence, this conversion determines the value of diffuse
flux at $\tau_V\sim 2$, which then propagates into the cloud interior
of $\tau_V\ga 2$.

A more rigorous analytic approach shows that the net flux is not
exactly zero. A very small flux, in comparison with the external flux,
goes through the cloud and creates a temperature gradient. This yields
the well known gray opacity solution \citep{Mihalas} where
$T^4(\tau)=a\tau+b$. While the constant $a$ depends only on the net
flux ($a=0$ in the case of zero net flux), the constant $b$ is
determined by the boundary condition on the cloud surface. Hence, this
becomes a reiteration of the role of the surface layer described
above.

The possibility of the temperature inversion effect described in this
paper was discussed previously by \citet{Wolf}. He noticed in his
numerical calculations that larger grains can have a higher
temperature than smaller grains in the surface of circumstellar dusty
disks. He correctly attributed this effect to the more efficient
heating of large grains by the IR radiation from the disk interior,
but did not analyze it further. This is the same effect noticed
by \citet{D02} in his numerical models of circumstellar disks with
gray dust opacities. However, it was not until \citet{Isella} and
\citet{VIJE06} described this effect analytically in more detail that
its importance to the overall disk structure was finally recognized.

The main driver for the detailed description of the temperature
inversion effect came from the need for a better understanding of
inner regions of dusty disks around young pre-main-sequence stars.
The advancements of near infrared interferometry enabled direct
imaging of these inner disk regions and resulted in the discovery of
inner disk holes produced by dust sublimation \citep{interfer}.
However, developing a self-consistent model that would incorporate the
spectral and interferometric data proved to be a difficult
problem. Two competing models are proposed. In one the data are
explained by a disk that has a large vertical expansion (puffing) at
the inner dust sublimation edge due to the direct stellar heating of
the disk interior \citep{DDN}. In the other model the inner disk is
surrounded by an optically thin dusty outflow, without a need for
special distortions to the vertical disk structure \citep{VIJE06}.

Although current observations cannot distinguish between these two models,
theory gives some limits on the dust properties in the former model.
The inner disk edge has to be populated by big dust grains ($\ga
1\mic$) \citep{Isella, VIJE06} in order to produce a large and
bright vertical disk puffing needed to fit the data. The hallmark
of this radiative transfer solution is the dust temperature
inversion, which we also demonstrate in \S\ref{single_grain}.
\citet{VIJE06} argue that purely big grains are
unrealistic because dust always comes as a mixture of grain sizes
(especially in dusty disks where dust collisions should constantly
keep small grains in the mix). On the other hand, \citet{Isella}
point out that smaller grains should be sublimated away from the
very surface of the inner disk edge, but cannot prove that this
process is efficient enough to preserve dust temperature
inversion.

Our analytical analysis in \S\ref{multigrain} proves that multigrain
radiative transfer solutions keep small grains away from the immediate
surface of optically thick disks and preserve temperature inversion.
This is achieved under the boundary condition of maximum possible
temperature reached by all dust grains. Then the surface becomes too
hot for small grains to survive, which leaves it populated only by big
grains. By choosing this favorable boundary condition, we made an
important presumption about the dust dynamics. Dust has to be
dynamically transported to the distances where all grains start to
sublimate.  This can be naturally occurring in optically thick disks
due to dust and gas accretion. However, in \S\ref{transverse} we
discovered that big grains can survive
closer to the star than the inner edge of optically thick disk. The
only requirement is that the disk becomes vertically optically thin at
these close distances.

\begin{figure}
 \epsscale{1.2}
 \plotone{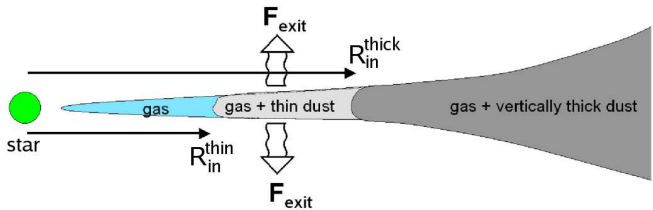}
 \caption{\label{Disksketch}
Dust dynamics in protoplanetary disks brings dust closer to the
star than the inner edge of optically thick dusty disk. According
to the radiative transfer solution in \S\ref{transverse}, such a
disk structure is possible if this zone is populated only by big
grains and it is vertically optically thin. These conditions
enable disk thermal cooling and dust survival (see also figure
\ref{Sketch3}). The resulting sublimation zone spans from the
optically thin to the optically thick inner disk radius (equation
\ref{Eq-Rin}). Its physical size is $\sim 0.2$AU in Herbig Ae stars
and $\sim 0.03$AU in T Tau stars.}
\end{figure}

Therefore, the dust sublimation zone is not a simple sharp step-like
transition, but rather a large zone (figure \ref{Disksketch}). We
estimate its size by considering the inner radius of optically
thick and thin disks \citep{VIJE06}
\eq{\label{Eq-Rin}
 R_{in} =  {\Psi R_*\over 2}\left({T_*\over T_{sub}}\right)^2
   = 0.0344 \Psi \left({1500K\over T_{sub}}\right)^2
      \sqrt{{L_*\over L_\odot}} \left[\hbox{AU}\right]
 }
where $T_*$, $R_*$ and $L_*$ are the stellar temperature, radius
and luminosity, respectively, and $\Psi$ is the the correction for
diffuse heating from the disk edge interior. Optically thick disks
have $\Psi=2$, while optically thin disks have $\Psi\sim 1.2$. The
sublimation zone exists between these two extremes, which
translates to $\sim 0.2$AU for Herbig Ae stars ($\sim 50L_\odot$)
or $\sim 0.03$AU in T Tau stars ($\sim L_\odot$). This is a
considerable size detectable by interferometric imaging and has to
be addressed in future studies. Also, its gas is enriched by
metals coming from the sublimated dust, which makes this zone a
perfect place for dust growth. Notice that, in addition to
sublimation, grain growth is another way of making the disk
optically thinner. Hence, the evolutionary role of such a large
sublimation zone has to be studied further in more detail.

Another major problem with the dynamics of big grains is that they
tend to settle toward the midplane, which suppresses vertical
expansion of the inner disk edge and leads to the failure of the disk
puffing model \citep{DD04,VIJE06}. On the other hand, the concurrent
model incorporating a dusty outflow lacks a convincing physical
process responsible for the formation of a dusty wind. The sublimation
zone may play an important role in dusty wind processes, since its small
optical depth and small distance from the star should also result in
gas properties that are more susceptible to non-gravitational forces
capable of launching a dusty wind (such as magnetic fields) than the
rest of the dusty disk.

\section{Conclusion}
\label{conclusion}

We analyzed the temperature structure of externally heated optically
thick dust clouds. We focused on the recently discovered effect of
temperature inversion within the optically thin surface of a cloud
populated by big ($\ga 1\mu$m) dust grains \citep{Isella, VIJE06}. The
effect is produced by reprocessing the external radiation and not by
any additional energy source.  The inversion manifests itself as a
temperature increase with optical depth, before it starts to decrease
once the external directional radiation is completely transformed into
the diffuse thermal flux. Since small grains do not show this effect,
the open question was whether small grains would manage to suppress
this effect when mixed with big grains.

We show analytically that small grains remove the temperature
inversion of big grains if the overall opacity is dominated by small
grains. However, this does not happen in situations where the cloud is
close enough to the external energy source for dust to start
sublimating. Small grains acquire a higher temperature than bigger
grains and sublimate away from the immediate cloud surface.  The exact
grain size composition of the surface depends on the amount of
external flux because different grain sizes sublimate at different
distances from the surface. These distances are smaller than in
optically thin clouds because bigger grains provide shielding to
smaller grains from direct external radiation. We show that the
temperature inversion is always preserved if small grains are removed
from only $\tau_V\sim 1$ of the immediate surface.

If the boundary condition requires all dust sizes to maximize
their temperature, then all small grains are removed from the
surface layer. They can exist only within the cloud interior
$exp(-\tau_V)\ll 1$ where the external radiation is completely
absorbed. Such a condition is expected in protoplanetary disks
where dust accretion moves dust toward the star. The inner disk
radius is then defined by the largest grains, no matter what the
overall grain size composition, because the largest grains survive
the closest to the star and dictate the surface radiative
transfer.

A new problem arises in that case. Since the temperature inversion
keeps the very surface of the cloud below the sublimation temperature,
its dust can move even closer to the star. We show that this creates
an optically thin dusty zone inside the dust destruction radius of an
optically thick disk (figure \ref{Disksketch}). Only big grains can
survive in this zone. We estimate that its size is large enough to be
detected by near IR interferometry. It consists of big grains and gas
enriched by metals from sublimated dust, hence favorable for grain
growth. This shows that the geometry and structure of inner disks
cannot be determined by simple ad hoc boundary conditions. It requires
self-consistent calculations of dust dynamics combined with radiative
transfer calculations and dust sublimation.

\acknowledgments

Supports by the NSF grant PHY-0503584 and the W.M. Keck Foundation are
gratefully acknowledged.

\newpage

\end{document}